\documentclass[twocolumn,amsmath,amssymb,aps,prb,superscriptaddress,nofootinbib]{revtex4-2}

\usepackage[english]{babel}
\usepackage[utf8]{inputenc}
\usepackage{xr-hyper}

\usepackage[colorlinks]{hyperref}
\usepackage[T1]{fontenc}
\usepackage[dvips]{graphicx}
\usepackage{amsmath}
\usepackage{amsfonts}
\usepackage{color}
\usepackage{ulem}
\usepackage[caption=false, justification=justified]{subfig}
\usepackage{url}
\usepackage{ulem}
\usepackage[top=2cm, bottom=2cm, left=2cm, right=2cm]{geometry}

\usepackage{soul}
\setstcolor{red}

\makeatletter
\newcommand*{\addFileDependency}[1]{
 \typeout{(#1)}
 \@addtofilelist{#1}
 \IfFileExists{#1}{}{\typeout{No file #1.}}
}
\makeatother

\newcommand*{\myexternaldocument}[1]{%
 \externaldocument{#1}%
 \addFileDependency{#1.tex}%
 \addFileDependency{#1.aux}%
}

\myexternaldocument{supplementary_information_PRL}

\newcommand{\degree}{$^{\circ}$}

\usepackage{xcolor}

\hypersetup{filecolor=blue}
\hypersetup{citecolor=blue}
\hypersetup{urlcolor=blue} 

\hypersetup{linkcolor=blue}

\begin{document}

\title{3D tomographic imaging of skyrmionic cocoons using HERALDO}

\author{Jhon~J.~Chiliquinga-Jacome}
\email{jhon.chiliquinga-jacome@cnrs-thales.fr}
\affiliation{Laboratoire Albert Fert, CNRS, Thales, Université Paris-Saclay, 91767, Palaiseau, France}

\author{Matthieu~Grelier}
\affiliation{Laboratoire Albert Fert, CNRS, Thales, Université Paris-Saclay, 91767, Palaiseau, France}
\affiliation{Present address: Spin-ion Technologies, 91120, Palaiseau, France}

\author{Riccardo~Battistelli}
\affiliation{Helmholtz-Zentrum Berlin, 14109 Berlin, Germany}
\affiliation{University of Augsburg, Augsburg, Germany}

\author{William~Bouckaert}
\affiliation{Laboratoire Albert Fert, CNRS, Thales, Université Paris-Saclay, 91767, Palaiseau, France}

\author{Krishnanjana~Puzhekadavil~Joy}
\affiliation{Helmholtz-Zentrum Berlin, 14109 Berlin, Germany}
\affiliation{University of Augsburg, Augsburg, Germany}

\author{Sophie~Collin}
\affiliation{Laboratoire Albert Fert, CNRS, Thales, Université Paris-Saclay, 91767, Palaiseau, France}

\author{Florian~Godel}
\affiliation{Laboratoire Albert Fert, CNRS, Thales, Université Paris-Saclay, 91767, Palaiseau, France}

\author{Marisel~Di Pietro Martínez}
\affiliation{Max Planck Institute for Chemical Physics of Solids, Dresden, Germany}
\affiliation{International Institute for Sustainability with Knotted Chiral Meta Matter (WPI-SKCM2), Hiroshima University, Hiroshima 739-8526, Japan}

\author{Claire~Donnelly}
\affiliation{Max Planck Institute for Chemical Physics of Solids, Dresden, Germany}
\affiliation{International Institute for Sustainability with Knotted Chiral Meta Matter (WPI-SKCM2), Hiroshima University, Hiroshima 739-8526, Japan}




\author{Felix~Büttner}
\affiliation{Helmholtz-Zentrum Berlin, 14109 Berlin, Germany}
\affiliation{University of Augsburg, Augsburg, Germany}

\author{Horia~Popescu}
\affiliation{Synchrotron SOLEIL, L’Orme des Merisiers, 91190, Saint Aubin, France}

\author{Vincent~Cros}
\affiliation{Laboratoire Albert Fert, CNRS, Thales, Université Paris-Saclay, 91767, Palaiseau, France}

\author{Nicolas~Reyren}
\email{nicolas.reyren@cnrs-thales.fr}
\affiliation{Laboratoire Albert Fert, CNRS, Thales, Université Paris-Saclay, 91767, Palaiseau, France}

\author{Nicolas~Jaouen}
\affiliation{Synchrotron SOLEIL, L’Orme des Merisiers, 91190, Saint Aubin, France}

\begin{abstract}
\textbf{Abstract:} 
Uncovering the rich and intricate characteristics of three-dimensional (3D) magnetic textures is essential for functional materials such as magnetic multilayers, where the delicate balance of various magnetic interactions leads to complex 3D spin arrangements. Among these textures, skyrmionic cocoons—tubular 3D magnetic structures characterized by a closed magnetization surface wrapping around a core—have emerged as particularly intriguing. Stabilized by competing magnetic interactions, these textures reside within a fraction of the thickness of the magnetic material and exhibit a typical lateral size of approximately 100 nm. Here, we present a vector tomographic reconstruction of the 3D magnetization in aperiodic Pt/Co/Al chiral multilayers, where skyrmionic cocoons have been recently reported. Using soft X-ray Holography with Extended Reference by Autocorrelation Linear Differential Operator (HERALDO), we acquire tomographic projections of the magnetic configuration and reconstruct the full 3D magnetization vector field with a spatial resolution of approximately 30 nm, as determined by Fourier shell correlation (FSC). This resolution allows us to observe critical features of the cocoons, such as their vertical misalignment and their overall chirality. Our findings demonstrate that HERALDO-based vector tomography is a powerful approach for revealing the internal structure and vertical extent of these nanoscale magnetic textures, offering new experimental insights into their intrinsic behavior.

\end{abstract}

\maketitle

\section{INTRODUCTION}

Due to the emergence of three-dimensional (3D) nanomagnetism as a next step for new generation spintronic devices~\cite{Zheng2018,Fernandez-Pacheco2017,Cheenikundil2022,Sahoo2021,BellizottiSouza2025}, a large interest for the understanding of 3D magnetic textures has arisen. Advances in synthesis methods and structural arrangement have led to the discovery of new unconventional non-collinear magnetic textures such as chiral bobbers~\cite{Kuchkin2025,Gong2021}, hopfions~\cite{Kent2021,Zheng2023} and skyrmionic cocoons. Skyrmionic cocoons are three-dimensional magnetic textures characterized by a closed surface of magnetization that wraps around a core, forming a tubular or "cocoon-like" structure. Internally, they exhibit a complex magnetization configuration, where the magnetic moments align in a non-collinear manner, creating a helical or swirling pattern along the axis of the cocoon. This internal structure is stabilized by the interplay of exchange interactions, Dzyaloshinskii-Moriya interactions (DMI), and magnetic anisotropy, giving rise to their unique 3D topology~\cite{Grelier2022,Grelier2023,Gopi2024}. Since the behavior of these textures is closely tied to their internal magnetization structures, a thorough and depth-resolved characterization is essential. To this aim, transmission-based methods have emerged as excellent powerful tools, as their imaging capabilities can be extended to obtain tomographic images of the magnetization configuration. For instance, neutrons~\cite{Henderson2023,Manke2010}, electrons~\cite{Silinga2025,Lyu2024} and x-rays~\cite{Donnelly2017,Donnelly2018,Seki2022,Hierro-Rodriguez2020,Hermosa-Munoz2022} have all been used for this purpose.

\medskip

In this context, soft x-ray techniques have been extensively used to investigate the magnetic structure of ferromagnetic thin films and multilayer systems leveraging the x-ray magnetic circular dichroism (XMCD). These techniques offer nanometer-scale spatial resolution, element-specific imaging and 
penetration depths of several hundreds of nm~\cite{Li2019,Butcher2025}. In particular, a well-established technique known as Fourier Transform Holography (FTH) has been used to image magnetic samples. This technique allows to obtain two-dimensional (2D) projections of the magnetic state with just one deterministic step~\cite{Eisebitt2004, Schaffert2013,Tieg2010,Zayko2021}. It involves coherently illuminating the object of interest along with a 
known reference, leading to interference between the diffracted light. The resulting diffracted pattern is recorded and the retrieved image is a convolution of the object and the reference, for which the spatial resolution is typically limited by the size and quality of the reference~\cite{Malm2022}. To enhance the image resolution, the FTH process can be further improved by phase retrieval algorithms~\cite{Marchesini2007}.

\medskip

Here we focus on Holography with Extended Reference by Autocorrelation Linear Differential Operator (HERALDO), with two orthogonal slits as references~\cite{Mustafi2023,Guizar-Sicairos2007}. This method enhances the reference wave, increases the total intensity of the reference contribution and, by tilting the sample around the slit normal, it allows the acquisition of off-normal projections of the sample's magnetization~\cite{Harrison2024,Kfir2017,Duckworth2013}. This approach enables tomographic imaging, providing access to features that are otherwise inaccessible through normal incidence, such as in-plane magnetization components ~\cite{Duckworth2011,Turnbull2021,Bukin2016,Parra2016}. Notably, we have recently shown that HERALDO can reveal the 3D nature of double skyrmionic cocoons by imaging tilted multilayer samples~\cite{Grelier2023_thesis}.

\medskip

In this study, we utilize HERALDO-based tomographic imaging with soft x-rays to achieve a full vectorial reconstruction of the magnetization in aperiodic magnetic multilayers, which host double skyrmionic cocoons, which reside in a fraction of the magnetic multilayers~\cite{Grelier2023}. Our primary objective is to investigate the internal configuration of these cocoons, building on a  previous work, where a 3D full-vectorial image of a Fe/Gd multilayer was obtained with 80-nm spatial resolution across a 5-$\mu$m-diameter field of view~\cite{DiPietro2023}. Here,  we apply HERALDO with two orthogonal slits to image a multilayer system hosting both double cocoons and 3D worm domains, the latter spanning across the full multilayer. In order to capture all three magnetization components, the sample is tilted along two perpendicular axes. We acquire 2D projections of the magnetic state at various tilt angles, confirming the existence of cocoons in the selected magnetic configuration. This observation is further validated by 2D images of the sample high-PMA part (middle layers of the stack) at normal incidence, which reveal no circular-shape objects, unlike those appearing when imaging the entire structure. To improve the resolution of the 3D reconstruction, we exclude from our analysis regions covered by the holography mask (object hole walls) during the sample rotation. 

\medskip

Our findings provide novel insights into the internal structure of skyrmionic cocoons. Our 3D reconstruction visually confirms the lateral non-alignment of cocoons, previously observed only in 2D projections~\cite{Grelier2022}, and offer a detailed visualization of their 3D morphology. The tomographic reconstruction reveals the spatial distribution of cocoons within the multilayer and their size variations across the sample, which we attribute to material grains and the complex fields generated by neighboring textures~\cite{Grelier2023_thesis}. These results highlight the sensitivity of cocoon formation to local material properties and magnetic environments. Additionally, our micromagnetic simulations, initialized with the reconstructed magnetization state, validate the overall structure of the cocoons and their domain walls, demonstrating the robustness of our imaging approach.

\medskip

While our reconstruction captures the essential features of the cocoons, the current spatial resolution of about 30 nm, as determined by Fourier Shell Correlation (FSC), limits our ability to resolve even finer details, such as layer-specific variations of the magnetic textures inside the cocoons. This highlights the need for further advancements in layer-resolved reconstruction techniques. Nevertheless, our findings clearly demonstrate that HERALDO-assisted tomographic 3D reconstruction of the magnetization using soft x-rays is a powerful and versatile tool for characterizing complex 3D magnetic textures. By providing direct access to depth-resolved information, this technique opens new paths for understanding and engineering magnetic nanostructures, paving the way for the next generation of spintronic devices.

\section{METHODS}
\label{Methods}

\subsection*{Sample preparation}

\begin{figure}
 \centering
 \includegraphics[width=0.48\textwidth]{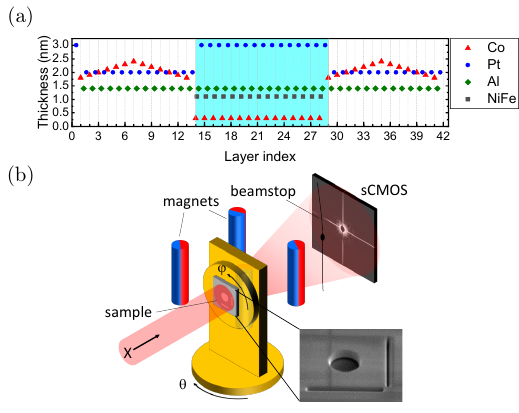}
\caption{(a) Thickness distribution for the materials that constitute the sample. The cyan background highlights the section with strong perpendicular magnetic anisotropy which contains NiFe as chemical marker for imaging. (b) Schematic configuration of the HERALDO measurement. Four permanent magnets with magnetization along a radial axis set the field at the sample position, using rotation and distance from the $\theta$ rotation axis. The front magnet is made transparent for clarity. The technical drawing is displayed in Fig.\,\ref{FigCOMET}. The zoom-in scanning electron microscopy image present a typical HERALDO mask seen from the back.}
\label{Fig1}
\end{figure}

Figure \ref{Fig1}a shows the composition of the sample used for the tomographic reconstruction consisting of three sections. The two outermost comprise a series of trilayers with varying Co thicknesses. These two ``gradient'' sections surround a central NiFe-based stack. The central one was designed to exhibit strong perpendicular magnetic anisotropy (PMA) by adding a thin Co layer below the NiFe one. Taking advantage of the element-specific sensitivity of the x-ray technique, we use the NiFe (Permalloy) layers at the center to discriminate 3D worms (penetrating the central permalloy section) and cocoon textures (absent in the central section). 
The full stack composition is as follows: Ta5|Pt3|(Co[1.8:0.1:2.4]|Al1.4|Pt2)|(Co [2.3:0.1:1.8]|Al1.4|Pt2)|(Co0.3|NiFe1.1|Al1.4|Pt3)$\times$15 (Co[1.8:0.1:2.4]|Al1.4|Pt2)|(Co[2.3:0.1:1.8]|Al1.4|Pt2). The notation (Co[$X_1$:$S$:$X_2$]|Al1.4|Pt2) is used as in ~\cite{Grelier2022}, which represents a transition in the Co thickness from $X_1$ to $X_2$ in thickness steps of $S$. All the thickness values are expressed in nm.

\medskip

The multilayer was grown on 200-nm Si$_3$Ni$_4$ membrane by room-temperature magnetron sputtering, similar to that presented in~\cite{Grelier2023}. The backside of the membrane was covered by an x-ray opaque mask made of Ta5|(Au25|Al2.5)$\times$40, where the thickness is expressed in nm, for a total thickness of 1.1\,$\mu$m. A 2.5-$\mu$m-radius object hole was milled in the mask down to the silicon membrane to allow x-ray transmission. As references, two 8-$\mu$m-long slits perpendicular to each other were milled through the entire sample. They are located 9\,$\mu$m away from the center of the object hole (Fig. \ref{Fig1}b). The object hole and the slits were made using focused ion beam (FIB) and their positions meet the separation condition required for the technique, preventing overlapping of the object-reference cross-correlation with the object autocorrelation in the reconstructions~\cite{Mustafi2023}.

\subsection*{HERALDO measurements}

The experiment was performed at the COMET end-station of the SEXTANTS beamline of the SOLEIL synchrotron~\cite{Popescu2019} at room temperature. The sample was illuminated with circularly polarized soft x-rays to probe the magnetization component parallel to the beam direction, and the resulting holograms were recorded using an sCMOS detector positioned 200 mm downstream of the sample, as shown in Fig.~\ref{Fig1}b. The detector features a 2048$\times$2048-pixel array with a physical pixel pitch of 11\,$\mu$m. A beamstop was used to block the direct beam and prevent degradation of the detector. The energy of the x-rays was set to 777.4\,eV to image the whole sample thickness (about 0.8\,eV lower than the Co L$_3$-Edge to increase transmission minimizing absorption and radiation damage), and to 852 eV to image the high PMA part (0.7 eV lower than the Ni L$_3$-Edge for similar reasons).

\medskip

We saturated the sample with positive out-of-plane field perpendicular to the sample surface (OOP), and then lowered its magnitude to 100\,mT. The resulting magnetic configuration was well-suited for 3D imaging, as it reveals worm-like domains going throughout the entire stack, along with cocoons located in the outermost parts of the double gradient. A total of 58 projections were acquired at different rotation angles. The maximum angle was limited by the thickness of the mask and the shape of the object hole. The rotation angle goes from $\theta=-45$\degree\;to $+45$\degree for one slit and from $\theta=-45$\degree\;to $+33$\degree\;for the other, 0\degree\;being normal incidence. XMCD projections were measured in 3\degree\;steps within the angular range specified. For each angle, the applied external field was kept perpendicular to the sample surface thanks to the four permanent magnets which provide a field that can be rotated and controlled in both direction and amplitude within the horizontal plane \cite{Popescu2019}. We ensure that the magnetic configuration remains the same during the rotation process by doing a full rotation from -45\degree\; to +45\degree\;and reconstructing one image at each end. The rotations were done in 1\degree\;step.

\medskip

The holograms were recorded up to the edge of the detector. In every angle the sample was illuminated with both 
circularly right and left polarized light. The total exposure time for each polarization was modified for each hologram, ranging from 67.5\,s to 814.5\,s, with 4500 frames in each measurement. This change was done to obtain similar intensity in each hologram given that it decreased at higher sample rotation angles.

\subsection*{Holograms processing and magnetization reconstruction}

The two-dimensional (2D) projections for each angle have been reconstructed 
following the protocol outlined in~\cite{DiPietro2023}, making use of the tomographic reconstruction algorithm described in~\cite{Donnelly2018}. By using the HERALDO method~\cite{Guizar-Sicairos2007} we recover a real space image ($\phi$) from the hologram measured in the far field with circularly right ($\phi_+$) and circularly left ($\phi_-$) polarized light. The obtained images are a projection of the magnetization along the propagation direction of the x-rays.  Each reconstructed image is then numerically focused through wavefront propagation~\cite{Malm2022,Guehrs2010}. Because these images are complex-valued, we selected the propagation value that maximized the contrast of the imaginary component Im\{$\phi_- - \phi_+$\}. 

\medskip


The pixel size for the 2D reconstruction was estimated by doing the ratio between the known diameter of the object hole and its number of pixels in the reconstructed image at normal incidence. The latter was measured considering the edge of the object hole using Im\{$\phi_- + \phi_+$\}. This process gave us a pixel size of 15.6\,nm. Then, each image was cropped using a rotating cylindrical mask aligned to the rotation angle of the sample. The mask had a radius of 60 pixels (936\,nm) and a height of 15 pixels (234\,nm approximately the sample height), so as to exclude the region covered by the gold mask during sample rotation.

\medskip

The 3D reconstruction of the vectorial magnetization $\Vec{m}$ has been performed using the PyCUDA library \textsc{MAGTOPY}~\cite{DiPietro2023,Magtopy2023} (following~\cite{Donnelly2018}) with a 200$\times$200$\times$200 voxels volume. The resulting volume containing all the magnetic information is a cylinder with a radius of 60 voxels and a height of 14 voxels. The spatial resolution for each magnetization component has been estimated using FSC, splitting the data set in two.

\subsection*{Micromagnetic simulations}

The micromagnetic simulations were carried out using the software package MUMAX3~\cite{Vansteenkiste2014} at zero temperature. The micromagnetic parameters and the general implementation to simulate the double gradient structure were taken from our previous study~\cite{Grelier2023}.
 The saturation magnetization is $M_s=1.2$\,MA\,m$^{-1}$, the surface anisotropy constant is $K_{u,s}=1.62$\,mJ\,m$^{-2}$, the surface DMI constant is $D_s=2.34$ pJ\,m$^{-1}$ and the exchange constant is $A=18$\,pJ\,m$^{-1}$. The geometry is defined as a cylinder (due to the experimental reconstruction data shape) with a 936\,nm radius, corresponding to 234 cells. The simulation cells are 4$\times$4$\times$2.13\,nm$^3$.

\section{RESULTS AND DISCUSSION}

\begin{figure}
 \centering
 \includegraphics[width=0.51\textwidth, page =1]{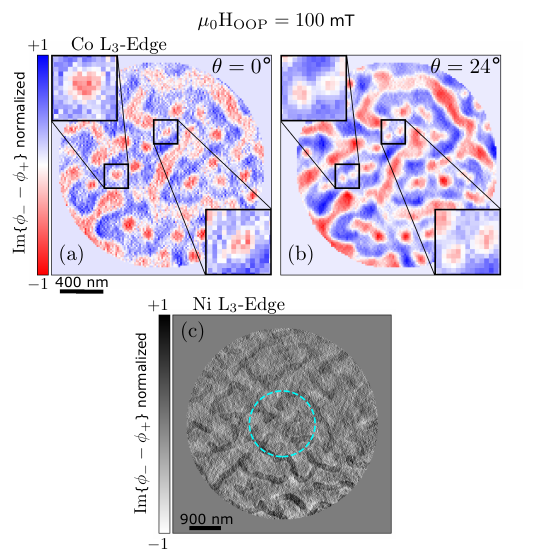}
 \caption{Normalized imaginary part of the HERALDO image reconstruction. (a) 0$^\circ$ and (b) 24$^\circ$ sample rotation angle and 100 mT OOP field at Co \mbox{L$_3$-Edge} (777.4\,eV). The insets correspond to a zoomed-in view of regions hosting magnetic cocoons. The same magnetic objects are highlighted for both rotation angles. (c) Image taken near the Ni \mbox{L$_3$-Edge} (852\,eV). The cyan circle surrounds the area shown in (a) and (b).}
\label{Fig2}  
\end{figure}

In Fig.~\ref{Fig2}, we present HERALDO image reconstructions for two representative rotation angles ($\theta=0^\circ$ and $\theta=24^\circ$). The images were acquired with a 100\,mT applied normal to the sample surface (OOP), enabling the investigation of  a magnetic configuration in which cocoons and 3D worms co-exist. The reconstruction was normalized for better comparison. The projections correspond to the 936-nm diameter disk (15.6 $\times$ 15.6\,nm$^2$ pixel size) that is visible for all rotation angles as discussed in section \ref{Methods}. No phase retrieval was used for the reconstructions.

\medskip

At normal incidence (see Fig. \ref{Fig2}a), several circular magnetic domains are observed, along with worm-like domains, both distinguished by dichroic contrast (red contrast against a blue background). In our previous work~\cite{Grelier2022},  we demonstrated that in a double gradient structure, these worm domains exhibit a columnar behavior, i.e., their magnetization is reversed in each magnetic layer relative to the background. 
By reconstructing an image at the Ni edge (see Fig.~\ref{Fig2}c), we confirm that in our system the worm domains are indeed going through all the structure because they were directly visible. In contrast, all the observed round features are cocoons, since they only appear on Co-edge images. Additionally, the similar contrast of these round objects indicates a shared internal magnetic structure. The insets show a zoom-in on two nearly round domains for closer examination.
 
\medskip
 
When the sample is tilted by $\theta=24^\circ$ as shown in Fig. \ref{Fig2}b, the main features of the worm domains retain their characteristic shape, as expected due to their columnar behavior. By contrast, some of the round textures are clearly split into two smaller ones. The insets show the same objects as in the panel (a). The two emerging features correspond to two individual cocoons, located at the outermost parts of the multilayer. Their slight vertical misalignment, rather than perfect stacking, causes the splitting to deviate from the horizontal axis. This misalignment also explains the apparent size difference between projections as the two cocoons projections partially overlap. This observation is likely due to pinning related to material grains~\cite{Grelier2023}. Interestingly, not all round objects in Fig. \ref{Fig2}a appeared as two when the sample is tilted, meaning that a higher angle is needed to separate them. This behavior may depend on their size and/or relative positions. In Fig.\ref{Fig2}b, some textures appear elongated, indicating partial appartent separation of the cocoons. While higher tilting angles generally result in a clear separation, distinguishing individual cocoons remains challenging if only relying in 2D projections, as their projections overlap with those of neighboring objects. To overcome this limitations we performed the 3D reconstruction of the magnetization vector, which provides depth resolution and eliminates the ambiguity caused by overlapping 2D projections. 

\begin{figure}
 \centering
 \includegraphics[width=0.47\textwidth]{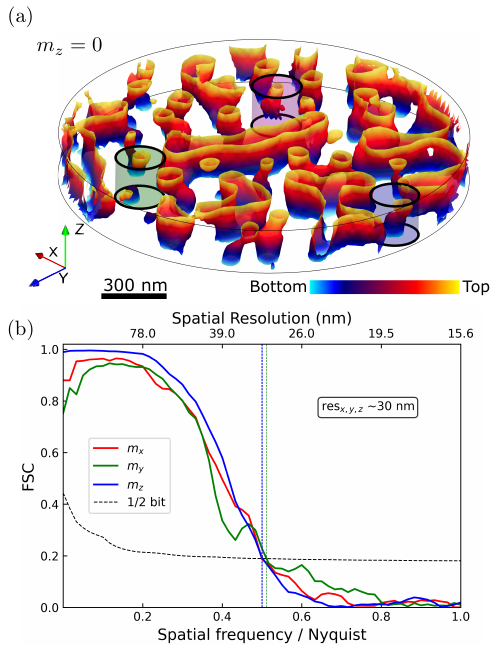}
\caption{(a) Three-dimensional imaging of the double gradient structure. The reconstructed region corresponds to a cylindrical volume of radius 60 voxels $\approx$ 936\,nm. The isosurfaces represent $m_z=0$, highlighting domain walls. The color gradient indicates the position of the corresponding feature in $z$. The sample end touching the substrate is displayed in cyan and the opposite end in yellow. The green volume inside features a double cocoon. The magenta and blue cylinders surround artificially joined cocoons. The $z$ dimension was oversampled from 13 to 50 voxels for better visualization, which does not change the features of the reconstruction. (b) Spatial resolution for each component of the magnetization estimated by Fourier shell correlation using the 1/2-bit criteria (dashed line).}
\label{Fig3}
\end{figure}

\medskip

After reconstructing and centering the projections for all rotation angles, the 3D image was computed using the python library MAGTOPY~\cite{DiPietro2023}. In Fig. \ref{Fig3}a, we show the 3D magnetic reconstruction for a 936-nm diameter disk of the double gradient structure. The isosurfaces represent $m_z=0$, emphasizing domain walls. The black lines show the disk's outline. 

\medskip

We observe that all the worm domains extend across the full thickness of the disk, in agreement with the reconstructions obtained at the Ni edge and earlier observations~\cite{Grelier2023}. However, some pair of cocoons, such as the one highlighted in green, do not extend through the central region. Additionally, several small objects are visible throughout the thickness, some of which exhibit a tilted structure (blue cylinder in Fig. \ref{Fig3}a), while others display a more complex behavior, including merging domains (magenta cylinder in Fig. \ref{Fig3}a). However, comparing to the projection taken at the Ni edge (Fig.~\ref{Fig2}c), we conclude that these features are artifacts of the reconstruction. The algorithm used for reconstruction can erroneously join textures or objects that appear merged or poorly separated in tilted projections due to the missing wedge in due to sampling or resolution problems.~\cite{Holler2019}

\medskip

The spatial resolution for each magnetization component has been estimated using the calculation of the FSC between two independent 3D reconstructions, splitting our images into two sets of 29 projections: (1) from $-45$\degree~to $+45$\degree~for one slit and from $-42$\degree~to $30$\degree~for the other with $6$\degree~steps; (2) from $-42$\degree~to $+42$\degree~for one slit and from $-45$\degree~to $33$\degree~for the other with $6$\degree~steps. A soft mask has been applied to each volume to reduce edge effects of the cylindrical volume. The calculation of the FSC was done using the python package PyNX~\cite{Favre-Nicolin2020}. As shown in Fig. \ref{Fig3}b, we use the 1/2-bit criteria~\cite{Vanheel2005} to determine the resolution value that corresponds to the crossing point of the FSC curves with the 1/2-bit threshold one (black dashed line). The calculated spatial resolution is approximately constant between $m_x$, $m_y$ and $m_z$, with a value of $\sim 30$ nm.

\medskip

Since the high PMA part of the double gradient structure is 87\,nm thick, this resolution should allow us to distinguish the vertical extent of the cocoons. Notably, the cocoons were consistently reconstructed in both independent analyses, which further validates the reconstruction process. It has to be noted that the effective resolution also depends on the quality of the two-dimensional reconstructed images. As seen in Fig. \ref{Fig3}a, although the resolution is in theory adequate, many cocoons appear as single object spanning through the whole thickness, revealing the limit of the FSC analysis (blue cylinder in Fig. \ref{Fig3}a). 
A higher resolution could be achieved by improving individual HERALDO images applying phase retrieval algorithms. A 5\,nm resolution was reported in Ref.~\cite{Battistelli2024} by using this approach. Additionally, the image quality is significantly impacted by signal loss due to the beamstop used (particularly at low frequencies). Recording the entirety of the hologram~\cite{Geilhufe2014} is expected to yield higher-quality 3D reconstruction.

\begin{figure*}
 \centering
 \includegraphics[width=0.8\textwidth]{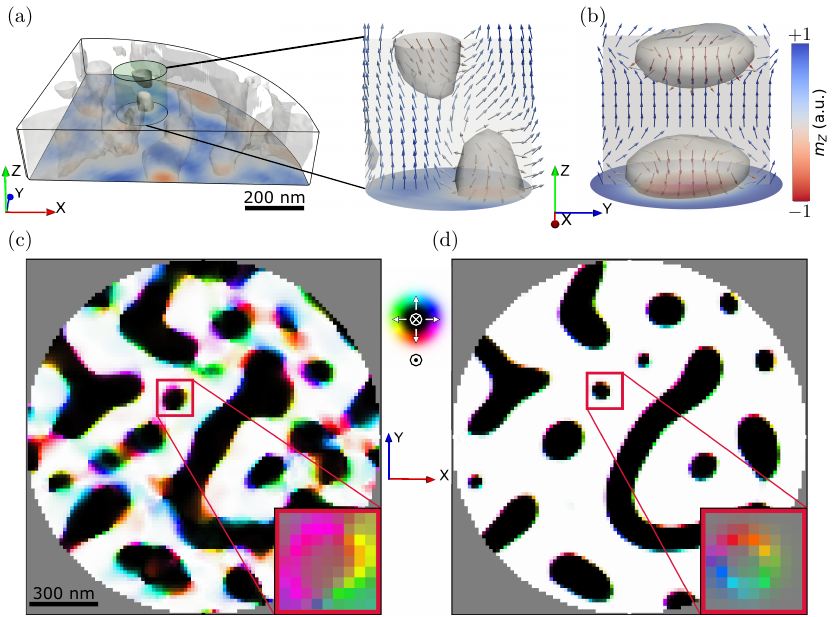}
 \caption{Comparison between three-dimensional tomographic reconstruction and relaxed micromagnetic simulation starting from the reconstructed state. The top row images show the isosurface $m_z=0$ in gray. A section of the reconstructed volume is shown in (a). The zoomed-in volume shows a double cocoon and its magnetization vector field for a given $yz$-plane. After relaxing the reconstructed state, the same region is shown in (b). The normalized sum of the magnetization vector field along $z$ for (c) the reconstructed and (d) relaxed state is depicted following the color wheel. The section of the reconstruction volume in (a) has an oversampling in the $z$ dimension from 14 to 50 voxels for better visualization while no resampling was applied for the zoomed-in region. For the cocoons in (a) and (c), the magnetization vectors are shown for the plane passing through the middle of the cocoons in both cases. The area shown in (c)-(d) corresponds to a circle with radius 780\,nm, lower than the simulated state to not consider the edge effects. The simulated state for (d) was subsampled after the relaxation process to match the in-plane pixel dimension of the reconstruction (d).}
\label{Fig4}
\end{figure*}

\medskip

To determine the angular resolution of the magnetization, we compare the measured magnetization vector field with a simulated one. The reconstructed magnetization configuration was used as an input into a MUMAX3~\cite{Vansteenkiste2014} micromagnetic simulation and relaxed under an OOP external field of 100\,mT, 
simulating each ferromagnetic layer (see Refs.~\cite{Grelier2022,Grelier2023}).
The resulting state was then resampled to match the in-plane dimensions of the reconstruction (120$\times$120 voxels) for better comparison.

\medskip

In Fig.~\ref{Fig4}a, a quarter of the total experimental reconstruction volume is shown, focusing on the region of the multilayer where a cocoon is present. A zoomed view of the cocoon on its right displays the magnetization vector field for a plane passing through the middle of the structure. We clearly see that the cocoons are not perfectly located one on top of each other, supporting our previous analysis regarding the 2D projections. In Fig.~\ref{Fig4}b, the simulated cocoons pair relaxed in the same region is shown with a resampling to 14 voxels in the $z$ direction. In the simulation both cocoons are vertically aligned because of attractive dipolar interaction which, in absence of pinning (e.g., crystalline grains), push the cocoons one on top of the other~\cite{Grelier2023}. Additionally, both cocoons have the same vertical extent, which is also the case for the reconstruction ones. Also, the size of different reconstructed cocoons varies along the sample, which can be attributed to material grains and the complex field generated by the neighboring textures which can change the structure of the cocoon~\cite{Grelier2023_thesis}. The reconstruction process introduces a smearing effect through the thickness, making it harder to clearly identify changes in ``effective'' chirality (i.e. sense of rotation of the in-plane component of magnetization) due to dipolar interactions, as observed in the simulation. The flux-closure is clearly observed in each cocoon in the relaxed state. In contrast, While the vertical resolution in the reconstruction is insufficient to reveal finer details as the abrupt change of chirality in the $z$ direction, the flux-closure behavior in each cocoon is still visible. The limited resolution plays an important role 
when analyzing the vertical change of the magnetization. Finally, 
numerical calculations \cite{Grelier2023} suggest that the cocoons configuration is not strongly affected by their misalignment, meaning that the observed chirality in the reconstruction is likely due to resolution artifacts.

\medskip

Fig. 4 displays the chirality of the textures, averaged over the entire material thickness. The in-plane component of the resulting vector field is represented by the colorwheel while its out-of-plane component is depicted through an illumination variation, as shown in the insets of Fig.~\ref{Fig4}c and~\ref{Fig4}d. Most of the textures present in the experimental reconstruction are also found in the relaxed simulation, indicating their stability.
In particular, we analyze in detail a cocoon texture in experimental images and simulations (insets in Fig.~\ref{Fig4}c and d). We first notice that their size is different, which can be attributed due to two contributions: First, the limited in-plane resolution of the reconstruction may widen the domain walls, making the objects appear larger in the normalized sum. Secondly, the misalignment of the textures in the reconstruction causes them to appear larger as seen in Fig.~\ref{Fig2}, which is not the case in the simulation. Finally, the reconstructed chirality of the chosen object aligns qualitatively well with the relaxed calculation. However, there are other objects where the chirality differs significantly from the simulated state. This discrepancy is more evident in regions where the reconstruction exhibits more noise, but it could correspond to existing topographical defects in the experimental system.

\section{CONCLUSIONS}

In conclusion, we show a three-dimensional tomographic reconstruction of the vectorial magnetization in thin-film multilayers hosting 3D magnetic textures: skyrmionic cocoons and worm-like magnetic domains. This was done by performing HERALDO with two orthogonal slit holographic references. The technique allowed us to sample the three orthogonal components of the magnetization. 
While FSC analysis estimates a resolution of about 30 nm consistent through all the magnetization components, the presence of artifacts discussed in the main text suggests that the actual value is lower.

\medskip

The reconstructed images confirm the 3D nature of the skyrmionic cocoons, which are clearly localized within part of the multilayered stack. The 3D reconstruction reveals also some tubular textures, the vertical extent of which needs probably further resolution, considering the estimated one. Notably, Ni-edge reconstructed holograms indicate that the reconstruction of these structures are affected by artifacts. The FSC approach does not account for all the resolution-related parameters of the reconstruction. Nevertheless, we could still resolve some of the 3D textures (cocoons). Additionally, we observe the influence of pinning effects, that in some cases prevent the vertical alignment of cocoons.

\medskip

To improve the reconstruction, higher sampling angles are needed, 
 although this would limit the effective field of view, which can be detrimental depending on the size of the magnetic textures analyzed. Additionally, the reconstruction algorithm did not account for varying thickness in the ferromagnetic material (for instance the gradient nature of two of the regions), which could affect phase modulation per layer and improve resolution if considered. Finally, removing the beamstop would allow the recovery of missing low-frequency information, resulting in more robust and quantitative reconstructions. \cite{Geilhufe2014,GEILHUFE2020}. We believe that our findings demonstrate that HERALDO-based vector tomography is a highly effective method for probing the internal structure and vertical scale of magnetic nanoscale textures, extracting crucial insights into their fundamental behavior directly from experimental data. 

\section{ACKOWLEDGEMENTS}

This study is supported by the French National Research Agency (ANR) and the German Research Foundation (DFG) through ANR-DFG project Topo3D (DFG grant no. 505818345), by the Horizon 2020 Framework Program of the European Commission under the EU project SkyANN (reference no. 101135729), by the Helmholtz Young Investigator Group Program through (grant no. VH-NG-1520), and by a France 2030 government grant managed by the French National Research Agency (grant no. ANR-22-EXSP-0002 PEPR SPIN CHIREX and grant no. ANR-22-EXSP-0008 PEPR SPIN SPINCHARAC). MDPM acknowledges the support of the French Embassy in Germany.

\section*{ANNEX}

\begin{figure*}
 \centering
 \includegraphics[width=1\textwidth]{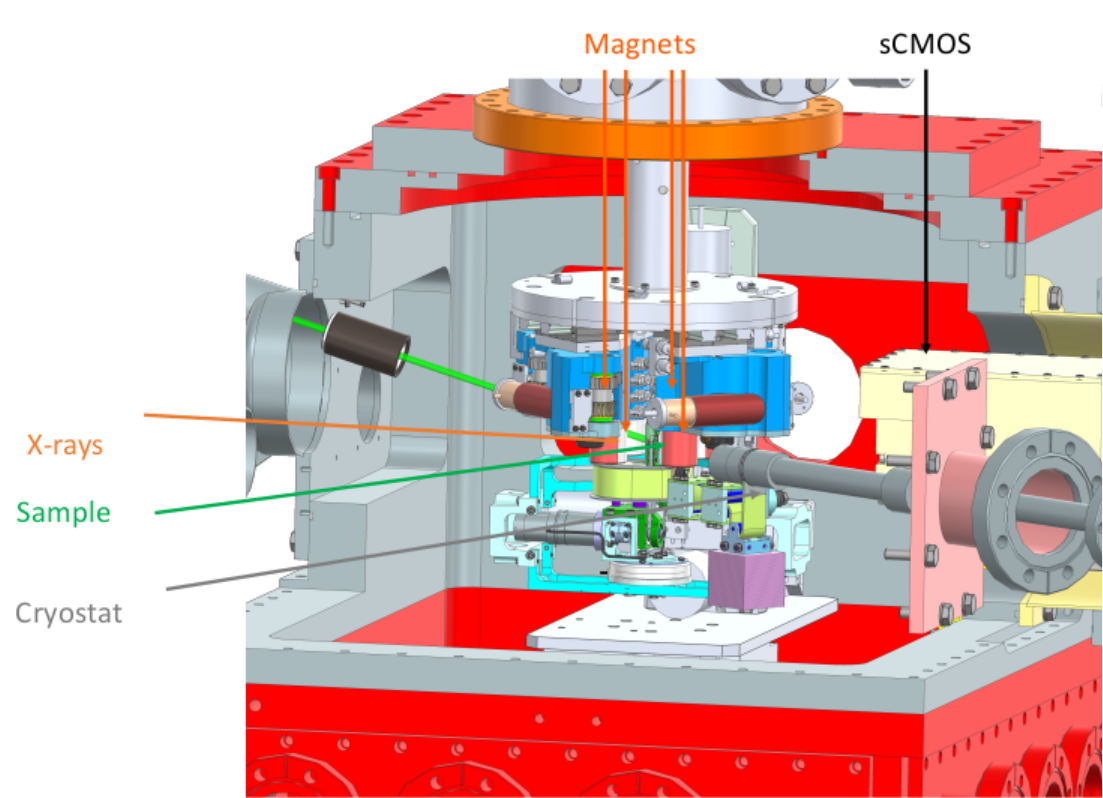}
\caption{Drawings of the COMET endstation in the 3D tomographic imaging configuration. X-rays arrive from the left, the sample holder is shown surrounded by the 4 permanents magnets (Cryostat not used in the experiment). The sCMOS detector is also highlighted.}
\label{FigCOMET}
\end{figure*}



\medskip

\bibliographystyle{ieeetr}
\bibliography{biblio}

\end{document}